\documentclass[10pt]{iopart}

\usepackage[english]{babel}
\usepackage{subfigure}
\usepackage{latexsym}
\usepackage{graphicx}
\usepackage{epsfig}

\begin{document}

\title{Variable stepsize Runge-Kutta methods for stochastic wave equations}
\author{Joshua Wilkie and Murat \c{C}etinba\c{s}}
\address{Department of Chemistry, Simon Fraser University, Burnaby, British Columbia V5A 1S6, Canada}
\begin{abstract}
We show that existing Runge-Kutta methods for ordinary differential 
equations (odes) can be modified to solve stochastic differential equations
(sdes) with strong solutions provided that appropriate changes are made to the way stepsizes are selected. The order of the resulting sde scheme is half the order of the ode scheme. Specifically, we show that an explicit 9th order Runge-Kutta method (with an embedded 8th order method) for odes yields an order 4.5 method for sdes which can be implemented with variable stepsizes. This method is tested by solving systems of sdes originating from stochastic wave equations arising from master equations and the many-body Schr\"{o}dinger equation. 
\end{abstract}

\maketitle 


\section{Introduction}

Stochastic wave equations play an important role in the quantum theory of decoherence and measurement\cite{GP,GAS} as well as in computational many-body physics\cite{CCD,JC,Wilk,Tess}. Solutions of master equations for completely positive dynamical semigroups\cite{CCD} and for Redfield theory\cite{GAS} can be expressed as expectations of diadics formed from wavefunctions obeying stochastic wave equations. Recently it has been shown that exact solutions of the N-body Schr\"{o}dinger or Liouville-von Neumann equations can be expressed as averages of Hartree products of single-body wavefunctions or densities which obey stochastic wave equations\cite{CCD,JC,Wilk,Tess}. Such methods could have important applications in chemistry and condensed matter physics. Stochastic differential equations (sdes) are also widely employed in other areas of physics, engineering and finance\cite{Gard}.

Unfortunately, efficient numerical techniques for solving such equations have not yet been developed. Algorithms in the literature have not substantially improved on the primitive methods described ten years ago in the well known text by Kloeden and Platen\cite{Platen}. Methods applicable to general systems of stochastic differential equations with multiple Wiener processes have not exceeded an order of 2. The low order of such methods restricts their domain of usefulness to one or few equations, and the larger systems of equations of interest in physics cannot be solved. 

Recently, one of us noted\cite{Wilk2} that with minor modifications classical methods for ordinary differential equations (odes) can be used to solve sdes with strong solutions. The technique was demonstrated by solving a wide range of low dimensional sdes with known exact solutions\cite{Wilk2}. Here we expand upon this idea by developing variable stepsize (i.e. adaptive) explicit Runge-Kutta based integrators for sdes. We demonstrate the use of the method by solving a variety of stochastic wave equations arising in decoherence problems\cite{GP}, and in stochastic decomposition of the many-body problem\cite{CCD,JC,Wilk,Tess}.

\section{Stochastic Taylor expansion}

There is a close connection between Taylor expansions of solutions of odes and Taylor expansions of strong solutions of sdes\cite{Wilk2}. Consider a finite set of sdes,
\begin{eqnarray}
dX^j_t=a^j({\bf X}_t,t)~dt+\sum_{k=1}^mb^j_k({\bf X}_t,t)~dW^k_t,
\label{sdes}
\end{eqnarray}
represented in It\^{o}\cite{Platen} form, where $j=1,\dots,n$. Here ${\bf X}_t=(X^1_t,\dots,X^n_t)$ and the $dW^k_t$ are independent and normally distributed stochastic differentials with zero mean and variance $dt$ (i.e. sampled $N(0,dt)$). The stochastic variables $W^k_t$ are Wiener processes. Assume that the coefficients $a^j$ and $b^j_k$ have regularity properties which guarantee strong solutions, i.e. that $X^j_t$ are some fixed functions of the Wiener processes, and that they are differentiable to high order. We may then view the solutions of (\ref{sdes}) as functions $X^j_t=X_j(t,W^1_t,\dots, W^m_t)$ of time and the Wiener processes. The solutions can therefore be expanded in Taylor series. Keeping terms of order $dt$ or less then gives
\begin{eqnarray}
X^j_{t+dt}&=&X^j_t+\frac{\partial X^j_t}{\partial t}~dt+\sum_{k=1}^m\frac{\partial X^j_t}{\partial W^k_t}~dW^k_t\nonumber \\
&+&\frac{1}{2}\sum_{k,l=1}^m\frac{\partial^2 X^j_t}{\partial W^k_t\partial W^l_t}~dW^k_tdW^l_t.\label{sdes2}
\end{eqnarray}
The product of {\em differentials} $dW^k_tdW^l_t$ is equivalent to $\delta_{k,l}dt$ in the It\^{o}\cite{Platen} formulation of stochastic calculus, so that
\begin{eqnarray}
dX^j_{t+dt}=X^j_{t+dt}-X^j_t&=&[\frac{\partial X^j_t}{\partial t}+\frac{1}{2}\sum_{k=1}^m\frac{\partial^2 X^j_t}{\partial W^{k2}_t}]~dt
\nonumber \\
&+&\sum_{k=1}^m\frac{\partial X^j_t}{\partial W^k_t}~dW^k_t.
\end{eqnarray}
Comparison to (\ref{sdes}) allows us to identify the first derivatives
\begin{eqnarray}
\frac{\partial X^j_t}{\partial W^k_t}&=&b^j_k({\bf X}_t,t) \\
\frac{\partial X^j_t}{\partial t}&=&a^j({\bf X}_t,t)-\frac{1}{2}\sum_{k=1}^m\frac{\partial^2 X^j_t}{\partial W^{k2}_t}\nonumber \\
&=&a^j({\bf X}_t,t)-\frac{1}{2}\sum_{k=1}^m\sum_{i=1}^nb^i_k({\bf X}_t,t)\frac{\partial b^j_k({\bf X}_t,t)}{\partial X_t^i}.\label{DERIVS}
\end{eqnarray}
From these first order derivatives, expressed in terms of $a^j$ and $b^j_k$, higher order derivatives can be computed. Thus a Taylor expansion of the solutions 
\begin{eqnarray}
X^j_{t+\Delta t}&=&X^j_t+\frac{\partial X^j_t}{\partial t}\Delta t+\sum_{k=1}^m\frac{\partial X^j_t}{\partial W^k_t}~\Delta W^k_t
\nonumber \\
&+&\frac{1}{2}\sum_{k,l=1}^m\frac{\partial^2 X^j_t}{\partial W^k_t\partial W^l_t}\Delta W^k_t\Delta W^l_t+\dots
\end{eqnarray}
can be obtained for finite displacements $\Delta t$ and $\Delta W^k_t$.

\section{Runge-Kutta methods for sdes}
 
This Taylor expansion of strong solutions of sdes can be employed to develop Runge-Kutta algorithms and other integration schemes\cite{Wilk2}. As an example consider the classic fourth order Runge-Kutta scheme with four stages
\begin{eqnarray}
K_j^1&=&f_j({\bf X}_{t_i},t_i)\nonumber \\
K_j^2&=&f_j({\bf X}_{t_i}+\frac{1}{2}{\bf K}^1,t_i+\frac{1}{2}\Delta t)\nonumber \\
K_j^3&=&f_j({\bf X}_{t_i}+\frac{1}{2}{\bf K}^2,t_i+\frac{1}{2}\Delta t)\nonumber \\
K_j^4&=&f_j({\bf X}_{t_i}+{\bf K}^3,t_{i+1})\nonumber\\
{\bf X}_{t_{i+1}}&=&{\bf X}_{t_i}+\frac{1}{6}({\bf K}^1+2{\bf K}^2+2{\bf K}^3+{\bf K}^4)
\label{rk4}
\end{eqnarray}
with $f_j({\bf X}_t,t)$ defined via
\begin{eqnarray}
f_j({\bf X}_t,t)&=&\frac{\partial X^j_t}{\partial t}\Delta t+\sum_{k=1}^m
\frac{\partial X^j_t}{\partial W^k_t}\Delta W^k_t\nonumber \\
&=&[a^j({\bf X}_t,t)-\frac{1}{2}\sum_{k=1}^m\sum_{i=1}^nb^i_k({\bf X}_t,t)\frac{\partial b^j_k({\bf X}_t,t)}{\partial X_t^i}]\Delta t
\nonumber \\
&+&\sum_{k=1}^mb^j_k({\bf X}_t,t) \Delta W^k_t.\label{RHS}
\end{eqnarray}
Here $t_i$ is the initial time and $t_{i+1}=t_i+\Delta t$. Taylor expansion shows that ${\bf X}_{t_{i+1}}$ differs from the exact solution by terms of order higher than $\Delta t^2$ (i.e. terms of higher order than $\Delta t^2$, $\Delta t (\Delta W^k_t)^2$, $(\Delta W^k_t)^4$, $(\Delta W^k_t)^2(\Delta W^l_t)^2$, and $(\Delta W^k_t)^2\Delta W^l_t\Delta W^i_t$). Thus, this stochastic Runge-Kutta algorithm is very similar to its classical counterpart except that its order is 2 not 4. 

Generalizations to higher order Runge-Kutta schemes are straightforward. One simply replaces the usual stage evaluations of Runge-Kutta with evaluations of (\ref{RHS}). Since (\ref{RHS}) is order $\Delta t^{1/2}$ rather than order $\Delta t$, the order of the sde method is half that of the ode method. SDE methods of order 2 and 4 constructed in this fashion have been shown to be very accurate in fixed stepsize calculations for small systems of sdes\cite{Wilk2}. In this manuscript we adapt a 9th order Runge-Kutta method\cite{Tsit} with 16 stages into an order 4.5 method for sdes.

However, fixed stepsize Runge-Kutta methods are neither accurate nor efficient for general systems of equations. To solve the large systems of sdes that arise in physical problems we need some means of controlling the local error.

\section{Adaptive stepsizes}

Local error is typically controlled in Runge-Kutta schemes for odes via the use of embedded lower order methods\cite{NR,Wann,DP1,DP2}. That is, Runge-Kutta methods can often be found wherein a method of order $l$ with $k\geq l$ stages has an embedded Runge-Kutta scheme of order $l-1$ which
uses some subset of the $k$ stages of the higher order method. Differences in the two solutions can be compared to a user requested tolerance to decide whether a contemplated step can be accepted or whether a smaller stepsize should be considered. Thus local error can be estimated, and stepsizes adapted to ensure the accuracy of the solution, at negligible extra cost. Well implemented examples of this approach are the 5(4) and 8(7) embedded pairs of Dormand and Prince (see \cite{DP1} and \cite{DP2}, respectively) which form the basis of the ode software package RKSUITE. A 9(8) pair has been derived by Tsitouras\cite{Tsit} although this algorithm has not been included in any software of which we are aware. Runge-Kutta methods of order 10 are known\cite{Hair} but embedded lower order pairs have not been reported. 

Variable stepsize one step schemes such as Runge-Kutta are popular because they are simple to understand and easy to implement. Multi-step schemes, such as Predictor-Corrector\cite{PC}, which store and use information from previous steps are however often much faster and more accurate. Unfortunately, it is not clear how the stochastic Taylor expansion developed above can be incorporated into a multi-step scheme. Predictor-Corrector\cite{PC}, for example, employs Lagrange type interpolation formulae (to fit $f_j({\bf X}_t,t)$ at a set of times), which are explicitly integrated over a time interval, to construct both the predictor and the corrector. It is 
far from clear how an analogous scheme would work for the $m+1$ variables $t,W_1,\dots, W_m$. Thus, Runge-Kutta methods seem to be the easiest to develop for sdes.

Once an error has been judged too large to be acceptable, ode codes simply try a smaller step and all information about the original step is lost. This procedure obviously cannot yield unbaised solutions in an analogous scheme for sdes. Therefore measures must be taken to ensure that the original Wiener process is maintained. One way of doing this is to halve the original step and to generate stochastic differentials on the two subintervals such that their sum is the original step. This approach was originally proposed by Gaines and Lyons\cite{GL} and is known as the method of binary Brownian trees. More sophisticated strategies have since been developed\cite{JRS} but they are specific to individual algorithms and cannot be easily adapted for our purposes. A number of schemes have been proposed for choosing the stochastic differentials on the subintervals\cite{GL,Lamba}. The correct approach appears to be that of Lamba\cite{Lamba} who generates the stochastic differential on the first subinterval by sampling the conditional probability
\begin{eqnarray}
\noindent p(\Delta W_a;\Delta W)&=&\frac{\int_{-\infty}^{\infty}dx_adx_b p(x_a)p(x_b)\delta (\Delta W_a-x_a) \delta (x_a+x_b-\Delta W)}{\int_{-\infty}^{\infty}dx_adx_b p(x_a)p(x_b)\delta (x_a+x_b-\Delta W)}\nonumber \\
&=&\frac{1}{\sqrt{2\pi (\Delta t/4)}}\exp\{-\frac{(\Delta W_a-\Delta W/2)^2}{2(\Delta t/4)}\}
\end{eqnarray}
where $\Delta W$ is the original stochastic differential, $\Delta W_a$ is the stochastic differential on the first subinterval and 
\begin{equation}
p(x)=\frac{1}{\sqrt{2(\Delta t/2)}}\exp\{-\frac{x^2}{2(\Delta t/2)}\}
\end{equation}
is the independent density for differentials on the subintervals.
This implies that the stochastic differential $\Delta W_a$ on the first subinterval has mean $\Delta W/2$ and variance $\Delta t/4$. The stochastic differential on the second subinterval $\Delta W_b$ must then be given by $\Delta W_b=\Delta W-\Delta W_a$ in order to maintain the original Wiener process. 

Thus, Runge-Kutta methods can be developed for sdes with strong solutions from Runge-Kutta methods for odes, and a binary tree variable stepsize strategy can be implemented with sampling on the subintervals via Lamba's method\cite{Lamba}. To show that the combined approach yields an accurate numerical method we solve a variety of stochastic wave equations from the recent physics literature. With the adaptive stepsize strategy we have chosen there is a good correlation between the speed of an algorithm and its order. We thus chose
to employ the highest order Runge-Kutta pair available, which as far as we are aware is the 9(8) pair of Tsitouras\cite{Tsit}. We have sucessfully used other lower order methods such as those of \cite{DP1} and \cite{DP2}, but for consistency all results reported in this paper were calculated using the method described in \cite{Tsit}.

\section{Examples}

Here we solve 3 sets of equations from the recent physics literature using the order 4.5 Runge-Kutta method implemented with variable stepsizes as described above.

\subsection{}

The first example we consider is the Gisin-Percival\cite{GP} stochastic wave equation for the nonlinear absorber 
(Eq. 4.2 of Ref. \cite{GP})
\begin{eqnarray}
d|\Psi(t,W_t)\rangle &=&.1(a^{\dag}-a)|\Psi(t,W_t)\rangle dt\nonumber \\
&+&(2\langle a^{\dag 2}\rangle a^2-a^{\dag 2}a^2-|\langle a^2\rangle|^2 )|\Psi(t,W_t)\rangle dt\nonumber \\
&+&\sqrt{2}(a^2-\langle a^2\rangle) |\Psi(t,W_t)\rangle dW_t\label{EG1}
\end{eqnarray}
where $a$ denotes the usual harmonic oscillator lowering operator. We chose an
initial state $|\Psi(0,0)\rangle=|0\rangle$ where $|0\rangle$ is the lowest eigenstate of $a^{\dag}a$. We choose $W_0=0$ in this and later examples and the Wiener process $W_t$ is real. The notation  $\langle Y\rangle $ indicates the quantum expectation $\langle \Psi|Y|\Psi\rangle$. The ensemble average over statistical realisations of the Wiener processes is denoted via $M[Y]$ for any $Y$. The quantity of interest for this example is the average density
\begin{equation}
\rho(t)=M[|\Psi(t,W_t)\rangle\langle \Psi(t,W_t)|]\label{DENS}
\end{equation}
which obeys the deterministic master equation
\begin{equation}
\frac{d\rho(t)}{dt}=.1[a^{\dag}-a,\rho(t)]+2a^2\rho(t)a^{\dag 2}-a^{\dag 2}a^2\rho(t)-\rho(t)a^{\dag 2}a^2.\label{ME1}
\end{equation}
\begin{figure}[htp]
\begin{center}
\epsfig{file=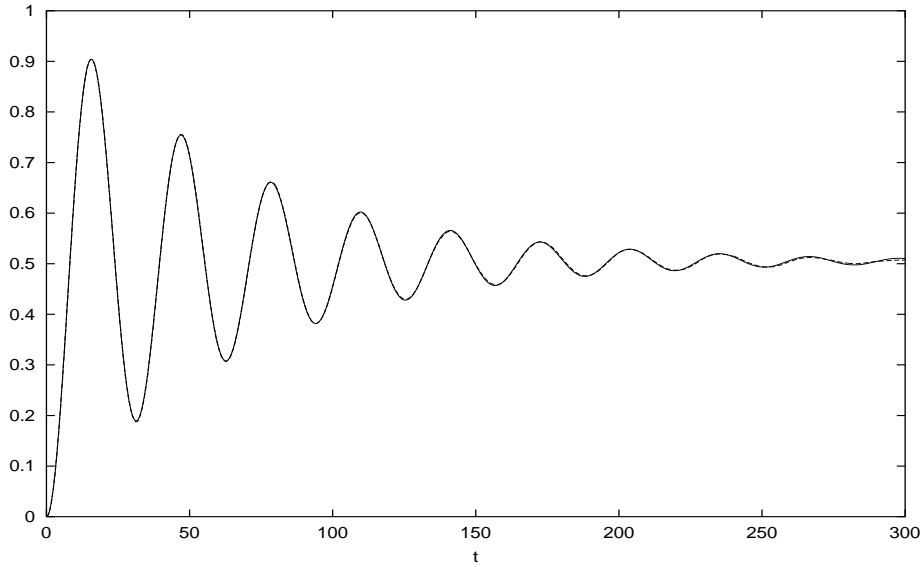,width=5.in,height=3.in}
\caption{Mean occupation number $n_t$ vs. time $t$}
\label{hesp}
\end{center}
\end{figure}

To implement our approach we need to find the derivatives of $|\Psi(t,W_t)\rangle$ with respect to $t$ and $W_t$. We immediately see that
\begin{equation}
\frac{\partial |\Psi(t,W_t)\rangle}{\partial W_t}=\sqrt{2}(a^2-\langle a^2\rangle ) |\Psi(t,W_t)\rangle
\end{equation}
and using (\ref{DERIVS}) we also determine that
\begin{eqnarray}
\frac{\partial |\Psi(t,W_t)\rangle}{\partial t}&=&.1(a^{\dag}-a)|\Psi(t,W_t)\rangle +(\langle a^4\rangle -\langle a^2\rangle ^2)|\Psi(t,W_t)\rangle \nonumber \\
&-&(a^2-\langle a^2\rangle )^2|\Psi(t,W_t)\rangle +(\langle a^{\dag 2}a^2\rangle -|\langle a^{2}\rangle |^2)|\Psi(t,W_t)\rangle \nonumber \\
&+&(2\langle a^{\dag 2}\rangle a^2-a^{\dag 2}a^2-|\langle a^2\rangle |^2)|\Psi(t,W_t)\rangle.
\end{eqnarray}
From these results we can now construct $f_j({\bf X}_t,t)$ using Eq. (\ref{RHS}). The Runge-Kutta scheme can thus be implemented as discussed above. The dynamics was solved in a basis consisting of the lowest 11 eigenstates $|n\rangle $ of $a^{\dag}a$ with $n=0,\dots, 10$. Thus, including real and imaginary parts of $\langle n|\Psi(t,W_t)\rangle$ for $n=0,\dots, 10$ our equations consist of a total of 22 real nonlinear coupled stochastic equations. A relative tolerance of $10^{-13}$ was requested. In Fig. 1 we plot the mean occupation number $n_t={\rm Tr}\{a^{\dag}a\rho(t) \}$ vs time for 50000 stochastic realisations (dashed curve) and for an exact solution of Eq. (\ref{ME1}) performed in the same basis set (solid curve). Agreement is very good.

Due to the increasing number of Wiener processes in the following examples, we drop the $W_t$'s as arguments of the 
stochastic wavefunctions. This change of notation is necessary but regretable since this dependence is an essential
requirement of the method we are testing.

\subsection{}

The second example is the Gisin-Percival\cite{GP} stochastic wave equation for a quantum cascade with absorption and stimulated emission (Eq. 4.4 of Ref. \cite{GP})
\begin{eqnarray}
d|\Psi(t)\rangle &=&-.1i(a^{\dag}+a)|\Psi(t)\rangle dt\nonumber \\
&+&(2\langle a^{\dag }a\rangle ~a^{\dag}a-(a^{\dag}a)^2-(\langle a^{\dag }a\rangle)^2)|\Psi(t)\rangle dt\nonumber \\
&+&.01(2\langle a^{\dag }\rangle a-a^{\dag}a-|\langle a\rangle |^2)|\Psi(t)\rangle dt\nonumber \\
&+&\sqrt{2}(a^{\dag}a-\langle a^{\dag}a\rangle ) |\Psi(t)\rangle dW_t^1\nonumber \\
&+&.1\sqrt{2}(a-\langle a\rangle ) |\Psi(t)\rangle dW_t^2.\label{EG2}
\end{eqnarray}
Here again we chose the initial state $|\Psi(0)\rangle =|0\rangle$. There are now 2 real statistically independent Wiener processes (i.e. $M[dW_t^1dW_t^2]=0$). The quantity of interest is again the density (\ref{DENS}) which in this case obeys the master equation
\begin{eqnarray}
\frac{d\rho(t)}{dt}&=&-.1i[a^{\dag}+a,\rho(t)]+2a^{\dag}a\rho(t)a^{\dag}a-(a^{\dag}a)^2\rho(t)-\rho(t)(a^{\dag}a)^2\nonumber \\
&+&.02a\rho(t)a^{\dag}-.01a^{\dag}a\rho(t)-.01\rho(t)a^{\dag}a.\label{ME2}
\end{eqnarray}
\begin{figure}[htp]
\begin{center}
\epsfig{file=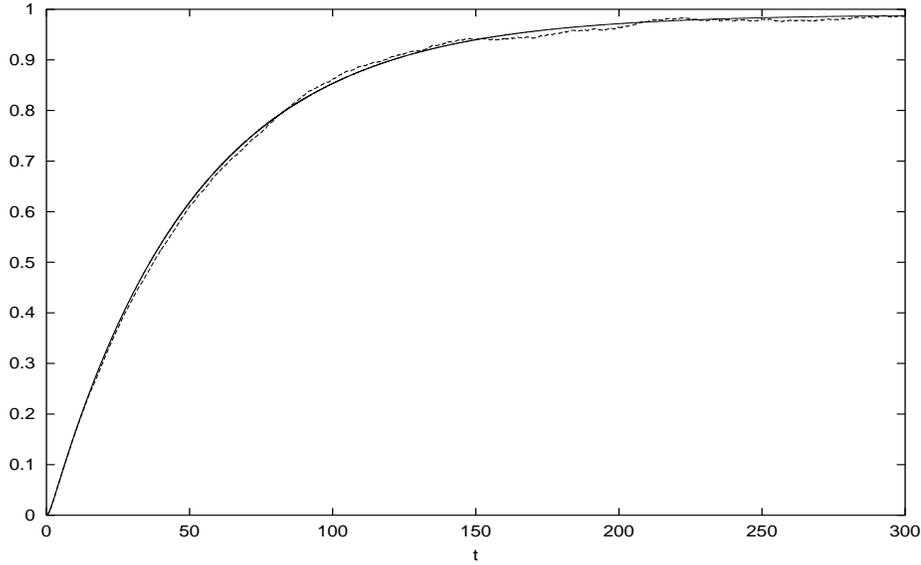,width=5.in,height=3.in}
\caption{Mean occupation number $n_t$ vs. time $t$}
\label{hesp2}
\end{center}
\end{figure}
The derivatives of $|\Psi(t)\rangle$ with respect to $t$, $W_t^1$ and $W_t^2$ are given by
\begin{eqnarray}
&&\frac{\partial |\Psi(t)\rangle}{\partial W_t^1}=\sqrt{2}(a^{\dag}a-\langle a^{\dag}a\rangle ) |\Psi(t)\rangle \\
&&\frac{\partial |\Psi(t)\rangle}{\partial W_t^2}=.1\sqrt{2}(a-\langle a\rangle ) |\Psi(t)\rangle \\
&&\frac{\partial |\Psi(t)\rangle}{\partial t}= -.1i(a^{\dag}+a)|\Psi(t)\rangle\nonumber \\
&+&(2\langle a^{\dag }a\rangle ~a^{\dag}a-(a^{\dag}a)^2-(\langle a^{\dag }a\rangle)^2)|\Psi(t)\rangle\nonumber \\
&+&.01(2\langle a^{\dag }\rangle a-a^{\dag}a-|\langle a\rangle |^2)|\Psi(t)\rangle \nonumber \\
&-&(a^{\dag}a-\langle a^{\dag}a\rangle)^2|\Psi(t)\rangle+2(\langle (a^{\dag}a)^2\rangle -\langle a^{\dag}a\rangle^2)|\Psi(t)\rangle\nonumber \\
&-&.01(a-\langle a\rangle )^2|\Psi(t)\rangle+.01(\langle a^{\dag}a)\rangle -|\langle a\rangle|^2)|\Psi(t)\rangle \nonumber \\
&+&.01(\langle a^2\rangle -\langle a\rangle^2)|\Psi(t)\rangle.
\end{eqnarray}
The same basis set and tolerance as in example 1 were employed.

Again we calculated the mean occupation number for 50000 trajectories (dashed curve) and for an exact solution of Eq. (\ref{ME2}) (solid curve). These quantities are plotted in Fig. (\ref{hesp2}). Agreement is again good but convergence is somewhat slower than in example 1 since we now have twice as many Wiener processes. 

\subsection{}

The third example consists of stochastic wave equations for a stochastic decomposition of the Schr\"{o}dinger equation 
for Helium\cite{Tess}. 

Neglecting nuclear motion about the center of mass, the Helium wavefunction $\Phi({\bf r}_1,{\bf r}_2,t)$ obeys the deterministic Schr\"{o}dinger equation (in atomic units $\hbar=1$, $m_e=1$, and $e=1$)
\begin{eqnarray}
\frac{\partial \Phi({\bf r}_1,{\bf r}_2,t)}{\partial t}&=&-i{\cal H}_2\Phi({\bf r}_1,{\bf r}_2,t)
\label{SE1}\\
&=&-i\{-\frac{1}{2}\nabla_1^2-\frac{1}{2}\nabla_2^2-\frac{2}{r_1} -\frac{2}{r_2}+\frac{1}{|{\bf r}_1-{\bf r}_2|}\}\Phi({\bf r}_1,{\bf r}_2,t),\nonumber
\end{eqnarray}
for any specified anti-symmetric initial state $\Phi({\bf r}_1,{\bf r}_2,0)$. Here ${\bf r}_1$ and ${\bf r}_2$ denote positions of electrons 1 and 2 with respect to the nucleus. For our example calculation we choose an initial wavefunction of the form
\begin{displaymath}
\Phi({\bf r}_1,{\bf r}_2,0) =
\beta \left( \Psi_{1}({\bf r}_1,0)\Psi_{2}({\bf r}_2,0)
- \Psi_{2}({\bf r}_1,0)\Psi_{1}({\bf r}_2,0)\right)\label{INIT}
\end{displaymath}
(where $\beta=1/\sqrt{2(1-|\langle \Psi_{1}(0)|\Psi_{2}(0)\rangle|^{2})}$ is a normalization factor) which is obviously antisymmetric in ${\bf r}_1$ and ${\bf r}_2$. Note that we are implicitly incorporating the two-component electron spins into the definitions of $\Psi_1$ and $\Psi_2$. For our purposes it is important that $\langle \Psi_{1}(0)|\Psi_{2}(0)\rangle\neq 0$. The actual initial conditions for this example calculation were chosen randomly as a mixture of 1s and 2s He$^+$ states for each electron. 

It can be shown\cite{Tess} that the exact deterministic wavefunction $\Phi({\bf r}_1,{\bf r}_2,t)$ evolving from (\ref{INIT}) can be decomposed into stochastic waves via an average of the form
\begin{eqnarray}
\Phi({\bf r}_1,{\bf r}_2,t)&=&\beta M[\Psi_{1}({\bf r}_1,t)\Psi_{2}({\bf r}_2,t)- \Psi_{2}({\bf r}_1,t)\Psi_{1}({\bf r}_2,t)]
\end{eqnarray}
where $\Psi_{1}$ and $\Psi_{2}$ satisfy stochastic wave equations
\begin{eqnarray}
&&d\Psi_{1}({\bf r},t) = 
[-i (-\frac{1}{2}\nabla^2-\frac{2}{r})\Psi_{1}({\bf r},t)-i\sum_{s=1}^p\omega_s\langle O_s \rangle_2 O_s \Psi_{1}({\bf r},t)\nonumber \\
&+&\frac{i}{2}\sum_{s=1}^p\omega_s \langle O_s \rangle_1 \langle O_s \rangle_2 \Psi_{1} ({\bf r},t) ] dt \nonumber \\
& + & 
\sum_{s=1}^p\sqrt{-i\omega_s} \left( O_s - \langle O_s \rangle_1 \right)
\Psi_{1}({\bf r},t) dW_t^s \\
&-&\sum_{s=1}^p|\omega_s|\frac{\langle \Psi_{1}|\Psi_{1}\rangle [\langle O_s^{\dag}O_s\rangle_1
- |\langle O_s \rangle_1|^{2}]}{2 {\rm Re} \left\{
\langle \Psi_{1} |\Psi_{2}\rangle \right\}}\Psi_2({\bf r},t)dt\nonumber \\
&&d\Psi_{2}({\bf r},t) =  
[-i (-\frac{1}{2}\nabla^2-\frac{2}{r})\Psi_{2}({\bf r},t)-i\sum_{s=1}^p\omega_s\langle O_s \rangle_1 O_s \Psi_{2}({\bf r},t)\nonumber \\
&+&\frac{i}{2}\sum_{s=1}^p\omega_s \langle O_s \rangle_2 \langle O_s \rangle_1 \Psi_{2} ({\bf r},t) ] dt \nonumber \\
& + & 
\sum_{s=1}^p\sqrt{-i\omega} \left( O_s - \langle O_s \rangle_2 \right)
\Psi_{2}({\bf r},t)  dW_t^s \\
&-&\sum_{s=1}^p|\omega_s|\frac{\langle \Psi_{2}|\Psi_{2}\rangle [\langle O_s^{\dag}O_s\rangle_2
- |\langle O_s \rangle_2|^{2}]}{2 {\rm Re} \left\{
\langle \Psi_{1} |\Psi_{2}\rangle \right\}}\Psi_1({\bf r},t)dt.\nonumber
\label{Oeqns2}
\end{eqnarray}
We have used a notation $\langle F \rangle_j=\langle \Psi_j|F|\Psi_j\rangle$ in the above equations. Here the $\omega_s$ and operators $O_s$ arise through the one-body expansion of the Coulomb interaction
\begin{equation}
\frac{1}{|{\bf r}_1-{\bf r}_2|}=\sum_{s=1}^p \omega_s O_s(1)O_s(2)
\end{equation}
which we performed numerically in a basis of He$^+$ eigenstates\cite{Tess}. In the calculation reported here $p=8$ which means that there are eight real Wiener processes. Since the initial states are of s type we included only the basis functions of s type with a principle He$^+$ quantum number of 4 or less\cite{Tess}. This means that the total number of equations was 32. Clearly this is by far the most computationally difficult of the three examples.

The derivatives of $\Psi_j({\bf r},t)$ are given by
\begin{eqnarray}
&&\frac{\partial \Psi_{j}({\bf r},t)}{\partial W_t^s}=\sqrt{-i\omega} \left( O_s - \langle O_s \rangle_j \right)
\Psi_{j}({\bf r},t) \nonumber \\
&&\frac{\partial \Psi_{j}({\bf r},t)}{\partial t}=
-i (-\frac{1}{2}\nabla^2-\frac{2}{r})\Psi_{j}({\bf r},t)-i\sum_{s=1}^p\omega_s\langle O_s \rangle_k O_s \Psi_{j}({\bf r},t)\nonumber \\
&+&\frac{i}{2}\sum_{s=1}^p\omega_s \langle O_s \rangle_1 \langle O_s \rangle_2 \Psi_{j} ({\bf r},t)  \nonumber \\
&-&\sum_{s=1}^p|\omega_s|\frac{\langle \Psi_{j}|\Psi_{j}\rangle [\langle O_s^{\dag}O_s\rangle_j
- |\langle O_s \rangle_j|^{2}]}{2 {\rm Re} \left\{
\langle \Psi_{1} |\Psi_{2}\rangle \right\}}\Psi_k({\bf r},t)\nonumber \\
&+&\frac{i}{2}\sum_{s=1}^p\omega_s(O_s^2-2\langle O_s \rangle_jO_s+2\langle O_s \rangle_j^2-\langle O_s^2 \rangle_j)\Psi_{j} ({\bf r},t)\nonumber \\
&+&\frac{1}{2}\sum_{s=1}^p|\omega_s|(\langle O_s^{\dag}O_s \rangle_j-|\langle O_s \rangle_j|^2)\Psi_{j} ({\bf r},t)
\end{eqnarray}
where $k\neq j$ and $j,k=1,2$. From these equations we can now construct $f_j({\bf X}_t,t)$ using Eq. (\ref{RHS}). 
\begin{figure}[htp]
\begin{center}
\epsfig{file=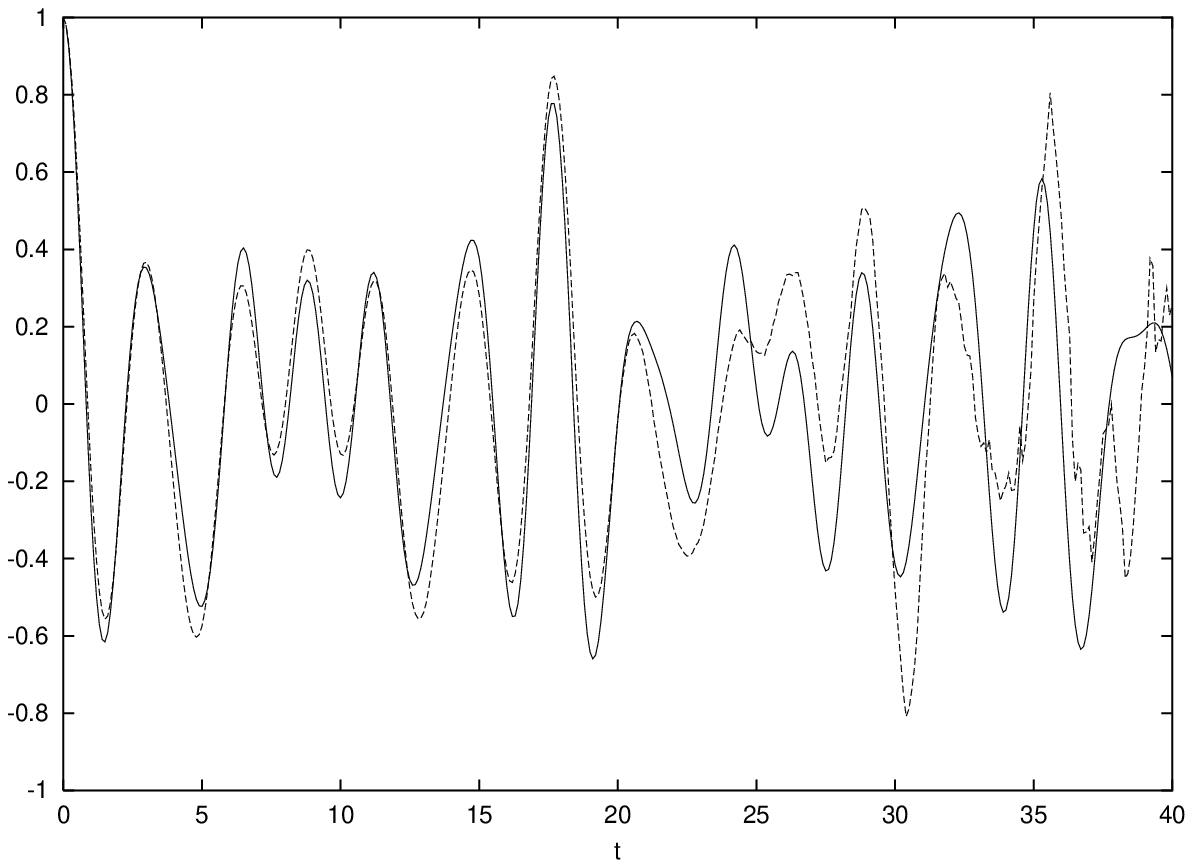,width=5.in,height=3.in}
\caption{Re $\langle \Psi(0)|\Psi(t)\rangle$ vs. $t$}
\label{repsi}
\end{center}
\end{figure}
\begin{figure}[htp]
\begin{center}
\epsfig{file=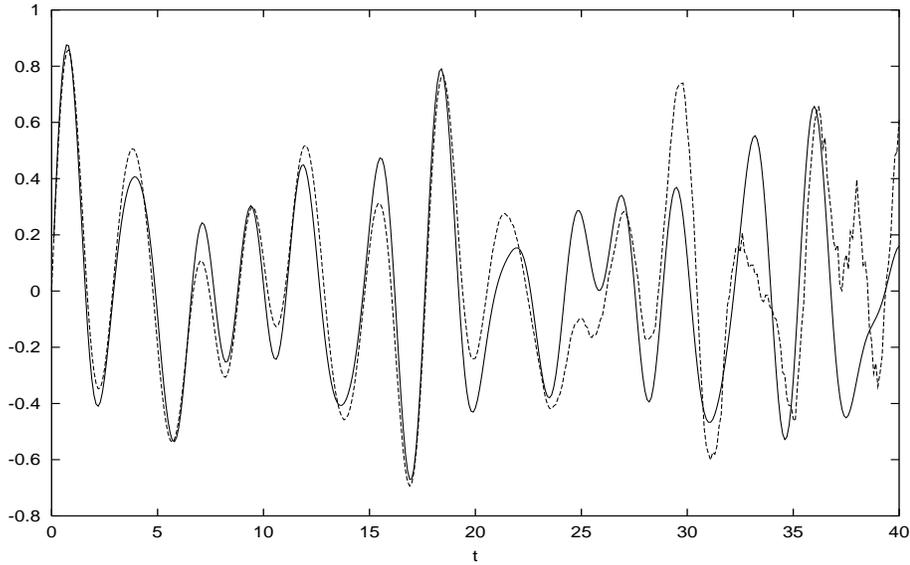,width=5.in,height=3.in}
\caption{Im $\langle \Psi(0)|\Psi(t)\rangle$ vs. $t$}
\label{impsi}
\end{center}
\end{figure}

A numerical problem arises in Eqs. (\ref{Oeqns2}) when the overlap $\langle \Psi_{1} |\Psi_{2}\rangle$ becomes small. The terms inversely proportional to this factor vary rapidly and the speed of integration slows down greatly. This occurs every few atomic units. Fortunately, this can be easily avoided by adding a small piece of $\Psi_{2}({\bf r},t)$ to $\Psi_{1}({\bf r},t)$, renormalizing the wavefunctions, and carrying the new norm as a weight factor in the stochastic average. The antisymmetric nature of the full wavefunction guarantees that this manipulation makes no change in the solution. 

In Figs. (\ref{repsi}) and (\ref{impsi}) we plot the real and imaginary parts of $\langle \Phi(0)|\Phi(t)\rangle$ for 200000 trajectories (dashed curve) and for the exact solution (solid curve) of the Shr\"{o}dinger equation (\ref{SE1}). Agreement is satisfactory with some deterioration of accuracy as time proceeds. 
\begin{figure}[htp]
\begin{center}
\epsfig{file=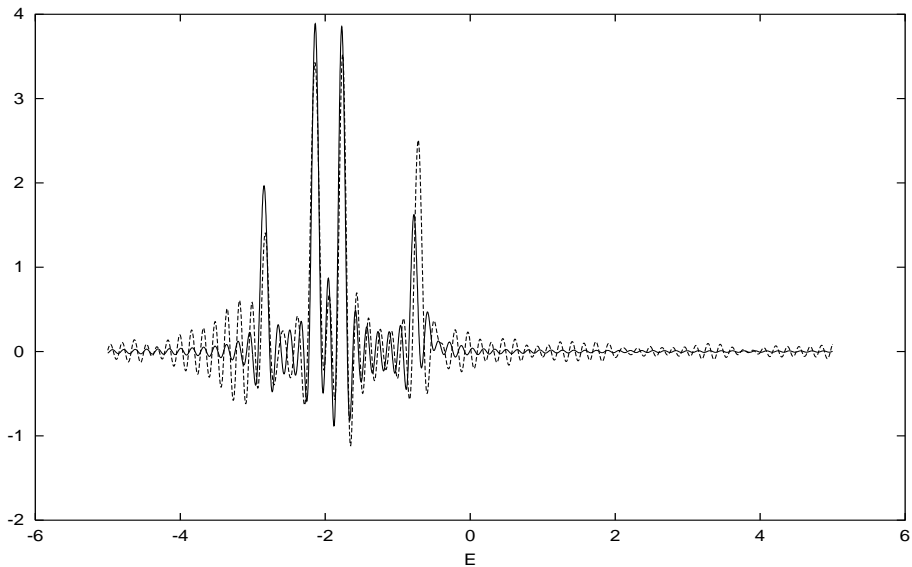,width=5.in,height=3.in}
\caption{He energy spectrum}
\label{hesp3}
\end{center}
\end{figure}

Finally, we computed the energy spectrum via
\begin{equation}
I(E) = \frac{1}{\pi\hbar} {\rm Re} \int_{0}^{T} \langle \Psi(0)|\Psi(t)\rangle
\exp \left(\frac{iEt}{\hbar} \right) \; dt
\simeq \langle \Psi(0)|\delta(E-{\cal H}_{2})|\Psi(0)\rangle
\label{spectrum}
\end{equation}
which we plot in Fig. (\ref{hesp3}). Again satisfactory agreement is obtained.

The fact that complete convergence is not achieved even with 200000 realisations may be due to the relatively large number of Wiener processes. Unfortunately, examples with very large numbers of Wiener processes will arise when the method of stochastic wave equations described in \cite{Tess} is applied to larger atoms or molecules. Thus it may be necessary to explore some form of importance sampling to improve convergence for these simulation methods. 

\section{Discussion}

The numerical strategy discussed in this manuscript provides a method for solving the large sets of coupled nonlinear sdes which arise in physical problems. This is currently the best strategy for solving systems of sdes like those that arise from stochastic wave equations. However, the large number of stages (16 for \cite{Tsit}) required for high order Runge-Kutta formulae limit the efficiency of our approach. Multistep methods for odes such as Predictor-Corrector\cite{PC} typically require only two evaluations of derivatives per step and can be implemented to any desired order. If such methods could be adapted for sdes the gain in efficiency could be enormous. Unfortunately, to implement such a strategy would require interpolation in $m+1$ variables with variable stepsizes (here $m$ is the number of Wiener processes). At present we do not see how this can be accomplished.

\ack{The authors acknowledge the support of the Natural Sciences and Engineering Research Council of Canada.}
~\\

\end{document}